\documentclass[preprint,12pt]{elsarticle}
 \usepackage{subfigure}
\usepackage{amsmath}
\usepackage{longtable,tabularx}
\usepackage{pgfplots}
\usepackage{calc}
\usepackage{textcomp}
\usepackage{float}
\usepackage{setspace}
\usepackage{natbib}
\usepackage{graphicx}
\usepackage{placeins}
\usepackage{caption}
\usepackage{array}
\usepackage{longtable}
\usepackage{multirow}
\usepackage{amssymb}
\pgfplotsset{width=7cm,compat=1.7}
\makeatletter
\def\fixedlabel#1#2{%
  \@bsphack%
  \protected@write\@auxout{}%
         {\string\newlabel{#1}{{#2}{\thepage}}}%
  \@esphack}
\makeatother
\journal{International Journal of Mechanical Sciences}
\DeclareUnicodeCharacter{2212}{-}
\begin{document}

\begin{frontmatter}



\title{Free Vibration analysis of Curvilinearly Stiffened Composite plates with an arbitrarily shaped cutout using Isogeometric Analysis}


\author[First]{Balakrishnan Devarajan\corref{cor1}}
\ead{dbalak9@vt.edu}

\cortext[cor1]{Corresponding Author}

\address[First]{Department of Biomedical Engineering and Mechanics, Virginia Polytechnic Institute and State University, Blacksburg, VA 24061, USA}

\begin{abstract}
This paper focuses on the isogeometric vibration analysis of curvilinearly stiffened
composite panels. The stiffness matrices and the mass matrices are derived using the first-order shear deformation theory (FSDT). The present method models the plate and the stiffener separately, which allows the stiffener element nodes to not coincide with the plate shell-element nodes. The stiffness and mass matrices of a stiffener are transformed to those of the plate through the displacement compatibility conditions at the plate–stiffener interface via the novel mapping technique. Cutouts are modeled using a single NURBS patch generated by creating a ruled surface between two curves.  The proposed formulation is first validated by comparing it with available literature. The effects of width-to-thickness ratio, fiber orientation, ply layups, shape and size of the cutouts and the boundary conditions on the response of stiffened composite plates are then analyzed and the numerical results are used to derive useful conclusions.
\end{abstract}



\end{frontmatter}


\section{Introduction}

Light-weight composite materials and structures have emerged as a cornerstone of modern engineering and manufacturing, offering unparalleled strength-to-weight ratios, design versatility, and energy efficiency across a wide range of applications, from aerospace to automotive industries \cite{patel2018lightweight, de2023biomass, braga2014advanced}. Due to their exceptional specific stiffness-to-weight and strength-to-weight ratios, composite materials are witnessing increased utilization in aircraft structures \cite{Renton2004}. Recent strides in manufacturing techniques, exemplified by the NASA Integrated Structural Assembly of Advanced Composites (ISSAC) robot and the creation of unitized stiffened composite panels through vacuum assisted resin transfer molding, have enabled the construction of intricate composite structures in atypical aircraft configurations, leading to lighter and more eco-friendly designs. These innovative manufacturing processes necessitate the development of efficient analysis and design tools tailored for stiffened composite panels produced using such methods.

The incorporation of curvilinear stiffeners, spars, or ribs in aircraft design introduces an expanded design domain, notably benefiting wing designs through heightened structural performance and reduced overall weight ~\cite{de2019structural, jrad2017global, robinson2016aeroelastic, de2021lightweight, de2017sparibs, de2019unconventional, de2017structural, de2018structural, de2020algorithms, demanual, kapaniamanual}. This is
due to the flexibility they offer in terms of geometric curvature in addition to the placement and orientation.

In practice, many aerospace structures encounter unpredictable pressure loads, making determinism impractical. Vibrations stemming from gust loads, engine noise, and auxiliary electrical systems can also generate noise within aircraft. Consequently, all flight vehicles function within a realm of random vibration. A comprehensive examination of finite element free vibration and dynamic analysis for composite laminated plates was presented by Zhang and Yang \cite{zhang2009recent}. Reddy and Khdeir \cite{reddy1989buckling} explored buckling and vibration responses in composite laminated plates, employing a variety of plate theories including classical, first, and third-order laminate theories. Carrera et al. \cite{carrera2012effects} scrutinized the vibration of slender composite plates under various in-plane loads, utilizing Carrera's unified formulation and finite element techniques.

Stiffened composite structures have found extensive application across engineering contexts, enhancing vibration response while concurrently reducing panel weight. Lee and Lee \cite{lee1995vibration} delved into the effects of fiber ply orientation, dimensions, and stiffener placement on free vibration modes of anisotropic stiffened plates, employing the first-order shear deformation theory. Rikards et al. \cite{rikards2001analysis} adopted an equivalent layer shell theory to examine buckling and vibration responses in laminated composite stiffened shells. Patel et al. \cite{patel2006buckling} investigated the static and dynamic instability traits of stiffened shells subjected to uniform in-plane harmonic edge loading using finite element methodologies.
Mukherjee and Mukhopadhyay \cite{mukhopadhyay1990finite} employed isoparametric shell elements to conduct vibrational analysis on plates with eccentric stiffeners. The displacement and geometry of the stiffeners were related to those of the plates using the displacement compatibility condition, which utilizes interpolation functions in the finite element method. Kumar and Mukhopadhyay \cite{kumar2000new} employed a beam element to represent the stiffener within a stiffened laminated plate. The displacement and geometry of the stiffener's beam element nodes were interpolated based on the nodes of the plate's shell element at the location of the stiffener's beam element node. Subsequently, this model was adopted by multiple researchers for both stiffened isotropic plates/shells and stiffened laminated plates in analyses involving buckling \cite{kumar2000new}, vibration \cite{shi2015vibration}, thermal effects, free vibration, and transient dynamic behavior \cite{Prusty2001}.

Isogeometric Analysis (IGA)  was introducted by Hughes et al. \cite{Hughes2005,cottrell2009isogeometric} which implements an isoparametric formulation by using NURBS basis functions to describe the geometry and to construct the finite basis approximations. The IGA developed rapidly and has been successfully applied in various fields \cite{cottrell2006isogeometric,manh2011isogeometric,bazilevs2012isogeometric,nguyen2015extended,tran2015vibration} including laminated composite plates \cite{kapoor2012geometrically,kapoor2013interlaminar,tran2014isogeometric} and functionally graded plates under a thermal environment \cite{tran2013isogeometric,phung2017nonlinear,carrera2005transverse}. Even though Qin et al.\cite{qin2017free} used IGA isogeometric analysis to analyze the static response of stiffened panels, studies emphasizing isogeometric analysis for the vibrational behvaior of stiffened composite plates are limited.

Due to an increase in requirement of laminated composite for a plethora of engineering applications,  the use of plates with arbitrarily shaped cutouts are unavoidable. The response of structures to loads can be significantly affected due to the presence of cutouts. Since the study of laminated composites with cutouts is a complex problem, numerical methods are used extensively. Vibration and buckling analysis of such structures have been performed using various numerical approaches like the finite strip method \cite{Eccher2008}, the finite element method (FEM) \cite{Chai1996,AydinKomur2010,Kumar2010}, meshfree methods \cite{Liu2008}, Rayleigh \textemdash Ritz method \cite{Chen2000} and \cite{Ovesy2012}, extended finite element method (XFEM) \cite{natarajan2014analysis}. Isogeometric analysis \cite{Hughes2005, Yu2015} is being widely used over the last ten years. It offers many benefits such as exact geometry representation, easy mesh refinement, higher-order continuity, and it avoids the mesh generation procedure when using the traditional way. Many insights into splines techniques \cite{Bazilevs2010}, mathematical properties \cite{Bazilevs2008, Evans2009a} and integration method \cite{Hughes2010a, Auricchio2012a} have been gained. Many problems have been successfully solved using IGA including fluid mechanics \cite{Bazilevs2008}, plates and shells \cite{Valizadeh2013, yin2014isogeometric, kapoor2013interlaminar, kapoor2012geometrically}   damage and fracture mechanics \cite{Verhoosel2011a},   contact mechanics \cite{Lu2011} and structural shape optimization \cite{Wall2008}.
Based on the best knowledge of the authors, no such identical task has been examined when this manuscript is being reported. The first section takes the readers through the mathematical formulation of the plate and the stiffener followed by the modeling technique used for generating arbitrary shaped cutouts. The final section discusses the results and the advantages of the present method over some existing methods.
\section{Mathematical Foundation}
\label{sec:sample1}
\begin{figure}[htbp]
  \centering
 \includegraphics[width=1\textwidth]{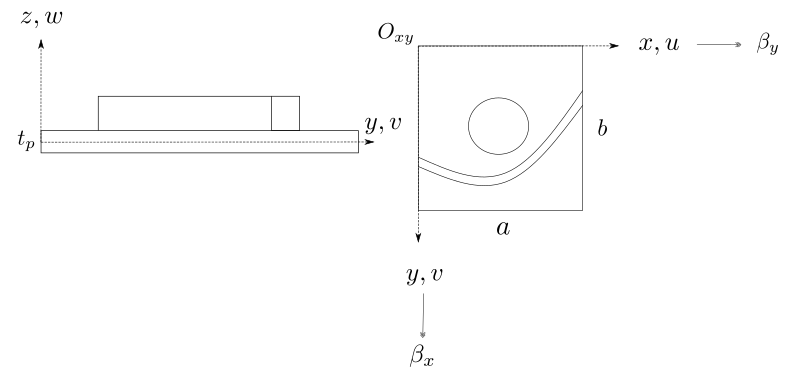}
  \caption{Geometry and nomenclature of a stiffened composite panel with a central cutout. Redrawn and modified from \cite{Zhao2016}}
\label{fig:vib_plate_with_a_cutout}
\end{figure}

In the first-order shear deformation theory (FSDT) the displacement field is
considered as the first-order Taylor expansion of mid-plane variables with respect
to plate thickness as follows:
Consider a stiffened composite panel as shown in Figure \ref{fig:vib_plate_with_a_cutout}.
The mid-plane of the panel $O_{xy}$ the global coordinate system. The plate is assumed to have a uniform thickness with no plydrops. Let $u$ be the displacements in the $x$-axis and $v$ be the in-plane displacements along the $y$-axis. Let $w$ be the transverse deflection along the $z$-axis.

\FloatBarrier  \FloatBarrier

Let $\beta_{y}$ and $\beta_{x}$ be rotation components of the panel around the $x$- and $y$-axes, respectively. The displacement components of the panel is defined by the first-order shear-deformable theory as follows,
\begin{equation}
\begin{array}{cc}&u\left(x\text{,}y\text{,}z\text{,}t\right)=u_0\left(x\text{,}y\text{,}z\text{,}t\right)+z\beta_x\left(x\text{,}y\text{,}t\right)\\&v\left(x\text{,}y\text{,}z\text{,}t\right)=v_0\left(x\text{,}y\text{,}z\text{,}t\right)+z\beta_y\left(x\text{,}y\text{,}t\right)\\&w\left(x\text{,}y\text{,}z\text{,}t\right)=w_0\left(x\text{,}y\text{,}z\text{,}t\right)\end{array}
\end{equation}

The plate strain energy $U_p$ can be written as,
\begin{equation}
    U_p=\frac12\underset\Omega{\hspace{0.28em}\int\int\hspace{0.28em}}\boldsymbol \varepsilon_\mathbf p^{\mathbf L\mathrm T}{\mathbf D}_\mathbf p\boldsymbol \varepsilon_\mathbf p^\mathbf L\mathrm d\Omega
\end{equation}

Where, $\mathbf D_{p}$ depends on the material property, stacking sequence, ply orientation and thickness of the plate. Derivation of $\mathbf D_{p}$  is explained in Section \ref{ss:mp_Orthotropic Layers}.
The generalized strains $\mathbf \varepsilon_{p}$  and the generalized displacements $\mathbf {u_{p}}$  of the panel can be written as,
\begin{equation}
\boldsymbol \varepsilon_\mathbf p^\mathbf L=
\begin{Bmatrix} 
\varepsilon_{x}^{0} \\ 
\varepsilon_{y}^{0}  \\ 
\gamma_{xy}^{0}  \\ 
\kappa_{x}^{0} \\
\kappa_{y}^{0} \\
\kappa_{xy}^{0} \\
\gamma_{xz}^{0} \\
\gamma_{yz}^{0} \\
\end{Bmatrix}=\begin{bmatrix} 
\frac{\partial}{\partial{x}} & 0 & 0 & 0 & 0 & 0 \\ 
0 & \frac{\partial}{\partial{x}} & 0 & 0 & 0 & 0\\ 
\frac{\partial}{\partial{x}} & \frac{\partial}{\partial{x}} & 0 & 0 & 0 & 0\\ 
0 & 0 & 0 & 0 & \frac{\partial}{\partial{x}} & 0\\ 
0 & 0 & 0 & 0 & 0 & \frac{\partial}{\partial{x}}\\ 
0 & 0 & 0 & 0 & \frac{\partial}{\partial{x}} & \frac{\partial}{\partial{x}}\\ 
0 & 0 & 0 & \frac{\partial}{\partial{x}} & 1 & 0 \\
0 & 0 & 0 & \frac{\partial}{\partial{x}} & 0 & 1 \\
\end{bmatrix}\begin{Bmatrix} 
u_{0} \\ 
v_{0}  \\ 
w_{0}  \\ 
\beta_{x} \\
\beta_{y}
\end{Bmatrix}=\mathbf {B_{p}} \mathbf {u_{p}} 
\end{equation}
\newline
The panel strain energy $U_{p}$ for the composite panel is,
\begin{equation}
U_{p}=\frac{1}{2}\iint_{\Omega}^{}
\mathbf{u_{p}}
^T
\mathbf{B_{p}}
^T
\mathbf{D_{p}}
\mathbf{B_{p}}
\mathbf{u_{p}}
d\Omega
\end{equation}

The kinetic energy $T_{p}$ for the composite panel is,
\begin{equation}
T_{p}=\frac{1}{2}\iint_{\Omega}^{}\dot{\mathbf{u}}_{p}^Tm_{p}\dot{\mathbf{u}}_{p}d\Omega
\end{equation}
Where,
\begin{equation}
    {\mathbf m}_\mathbf p=\rho{\left(\begin{array}{ccccc}t&0&0&0&0\\0&t&0&0&0\\0&0&t&0&0\\0&0&0&t^2/12&0\\0&0&0&0&t^2/12\end{array}\right)}
\end{equation}
\subsection{Orthotropic layers}
\label{ss:mp_Orthotropic Layers}
The constitutive equation in the local coordinate system for the $k^{th}$ orthotropic elastic laminate can be derived from Hooke\textquotesingle s law as
\begin{equation}
\begin{Bmatrix}\sigma_1\\\sigma_2\\\tau_{12}\\\sigma_3\\\tau_{31}\\\tau_{23}\end{Bmatrix}^{(k)}=\begin{bmatrix}Q_{11}&Q_{12}&0&0&0&0\\Q_{21}&Q_{22}&0&0&0&0\\0&0&Q_{66}&0&0&0\\Q_{31}&Q_{32}&0&Q_{33}&0&0\\0&0&0&0&Q_{55}&0\\0&0&0&0&0&Q_{44}\end{bmatrix}^{(k)}\begin{Bmatrix}\varepsilon_1-\alpha_1\mathrm\Delta T\\\varepsilon_2-\alpha_2\mathrm\Delta T\\\gamma_{12}\\\varepsilon_3-\alpha_3\mathrm\Delta T\\\gamma_{31}\\\gamma_{23}\end{Bmatrix}^{(k)}
\end{equation}
where $Q_{ij}$  are the elastic coefficients in the material coordinate system,  $T$ is the temperature change and $\alpha_i$  is the thermal coefficient of expansion in the principal $i^{th}$-direction.
The  transformed material constants are expressed as,
\begin{equation}
\begin{Bmatrix}{\overline Q}_{11}\\{\overline Q}_{12}\\{\overline Q}_{22}\\{\overline Q}_{16}\\{\overline Q}_{26}\\{\overline Q}_{66}\end{Bmatrix}=\begin{bmatrix}c^4&2c^2s^2&s^4&4c^2s^2\\c^2s^2&c^4+s^4&c^2s^2&-4c^2s^2\\s^4&2c^2s^2&c^4&4c^2s^2\\c^3s&\mathit{cs}(s^2-c^2)&-\mathit{cs}^3&2\mathit{cs}(s^2-c^2)\\\mathit{cs}^3&\mathit{cs}(c^2-s^2)&-c^3s&2\mathit{cs}(c^2-s^2)\\c^2s^2&-2c^2s^2&c^2s^2&{(c^2-s^2)}^2\end{bmatrix}\begin{Bmatrix}Q_{11}\\Q_{12}\\Q_{22}\\Q_{66}\end{Bmatrix}\text{,}
\end{equation}
\begin{equation}
\begin{bmatrix}{\overline Q}_{44}&{\overline Q}_{13}\\{\overline Q}_{45}&-{\overline Q}_{36}\\{\overline Q}_{55}&{\overline Q}_{23}\end{bmatrix}=\begin{bmatrix}c^2&s^2\\-\mathit{cs}&\mathit{cs}\\s^2&c^2\end{bmatrix}\begin{bmatrix}Q_{44}&Q_{13}\\Q_{55}&Q_{23}\end{bmatrix}\text{,}\hspace{0.35em}{\overline Q}_{33}=Q_{33}
\end{equation}
\begin{equation}
\begin{Bmatrix}\alpha_x\\\alpha_y\\\alpha_\mathit{xy}\end{Bmatrix}=\begin{bmatrix}c^2&s^2\\s^2&c^2\\2\mathit{cs}&-2\mathit{cs}\end{bmatrix}\begin{Bmatrix}\alpha_1\\\alpha_2\end{Bmatrix}\text{,}\hspace{0.35em}\alpha_z=\alpha_3
\end{equation}
where $c= \cos\theta$ and $s= \sin\theta$   .
\newline Using this, $\mathbf{D}_p$ can now be written as 

\begin{equation}
\mathbf {D}_p=\begin{bmatrix}\mathbf A&\mathbf B&\mathbf 0\\\mathbf B&\mathbf D&\mathbf 0\\\mathbf 0&\mathbf 0&\mathbf A^\text{s}\end{bmatrix}
\end{equation}
where,
\begin{equation}
(A_\mathit{ij}\text{,}{B}_\mathit{ij}\text{,}D_\mathit{ij})=\int_{-h/2}^{h/2}\overline Q_\mathit{ij}(1\text{,}z\text{,}z^2)\text{d}z\text{,}\hspace{1em}A_\mathit{ij}^\text{s}=K\int_{-h/2}^{h/2}\overline Q_\mathit{ij}\hspace{0.12em}\text{d}z
\end{equation}
where $A_\mathit{ij}$, $B_\mathit{ij}$ and $D_\mathit{ij}$ are valid for $i, j = 1, 2, 6$, and $A_\mathit{ij}^\text{s}$ for $i, j = 4$, 5 according to the Voigt notation. $K$ denotes the transverse shear correction coefficient. A value of $K=5/6$ was used for the analyses.

\subsection{Curvilinear stiffener}
\label{ss:stiffmodel}
Consider a stiffener attached to a
plate as shown in the figure below.
\FloatBarrier \begin{figure}[htbp]
\centering
   \includegraphics[width=1\textwidth]{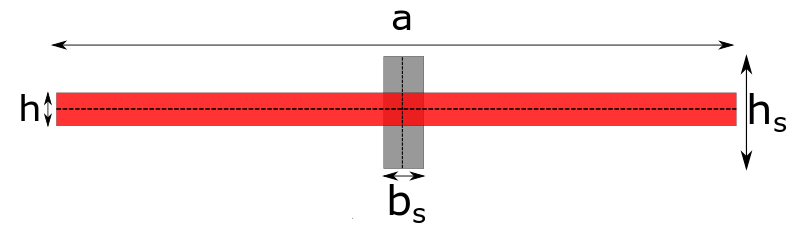}
  \caption{Composite plate (red) stiffened by a rectangular stiffener (gray)}
         \label{fig:mp_stiffener}
\end{figure} \FloatBarrier
The width and height of the rectangular stiffener are given as $b_s$ and $h_s$, respectively. Details of the coordinate system and nomenclature can be found in \cite{Zhao2016}. The stiffener is modeled
using 1D $2^{nd}$ order NURBS basis. The stiffener is modeled using the Timoshenko Beam theory.
The modeling of stiffener for free vibration analysis follows the same procedure as Section \ref{ss:stiffmodel}.
The stiffener strain energy $U_s$ is written as:

\begin{equation}
U_s=\frac12\int_{\Gamma}\mathbf \varepsilon_\mathbf s^{\mathbf L\mathrm T}{\mathbf D}_\mathbf s\mathbf \varepsilon_\mathbf s^\mathbf L\mathrm d\Gamma
\label{eq:birds11}
\end{equation}

The strain displacement relation of the composite stiffener
is \cite{Martini1988},
\begin{equation}
\mathbf{\varepsilon_{s}} =
\begin{Bmatrix} 
\varepsilon_{t}^{0} \\ 
\gamma_{n}^{0}  \\ 
\gamma_{b}^{0}  \\ 
\kappa_{t}^{0} \\
\kappa_{n}^{0} \\

\end{Bmatrix}=\begin{bmatrix} 
\frac{d}{dt} & \frac{1}{R} & 0 & 0 & 0 \\ 
 -\frac{1}{R} & \frac{d}{dt} & 0 & 0 & 0 \\ 
0 & 0 & \frac{d}{dt} & 1 & 0\\ 
0 & 0 & 0  & \frac{d}{dt} &\frac{1}{R}\\ 
0 & 0 & 0  & -\frac{1}{R} & \frac{d}{dt}\\ 

\end{bmatrix}\begin{Bmatrix} 
u_{t} \\ 
v_{n}  \\ 
w_{b}  \\ 
\beta_{t} \\
\beta_{n}
\end{Bmatrix}=\mathbf{B_{s}}\mathbf{u_{s}} 
\label{eq:birds21}
\end{equation}
where $\frac{1}{R}$ is the geometric curvature of the stiffener, computed at the integration points. Substituting  Equation \ref{eq:birds21} into Equation \ref{eq:birds11} we get,
\begin{equation}
U_s=\frac12\int_{\Gamma}\mathbf u_\mathbf s^\mathrm T\mathbf B_\mathbf s^{\mathrm T}{\mathbf D}_\mathbf s\mathbf B_\mathbf s^\mathbf L{\mathbf u}_\mathbf s\mathrm d\Gamma,
\end{equation}
The ${\boldsymbol D}_s$ in the above strain energy equation is the rigidity matrix for the composite stiffener and is given by,

\begin{equation}
  {\left(\begin{array}{ccccc}E_sA&0&0&E_sAe&0\\0&G_sA_n&0&0&G_sA_ne\\0&0&G_sA_be&0&0\\E_sAe&0&0&E_sI_n&0\\0&G_sA_ne&0&0&G_sJ_t\end{array}\right)}
\end{equation}

The kinetic energy for the stiffener $T_{s}$ is
\begin{equation}
T_{s}=\frac{1}{2}\int_{\Gamma}^{}\dot{\mathbf{u}}_{s}^Tm_{s}\dot{\mathbf{u}}_{s}d\Gamma
\end{equation}
where $m_s$ is the stiffener mass matrix and is given as,

\begin{equation}
m_{s}=\rho_{s}
\begin{bmatrix} 
A & 0 & 0 & Ae & 0 \\ 
0 & A & 0 & 0 & Ae \\ 
0 & 0 & A & 0 & 0\\ 
Ae & 0 & 0  & I_{n} & 0\\ 
0 & Ae & 0  & 0 & I_{n}+I_{b}\\ 

\end{bmatrix}
\end{equation}

Before building the linear system of equation of the stiffened plate, the degrees of freedom of the plate and curvilinear stiffener should be unified. Detailed steps to achieve this are as follows : 

\begin{itemize}
    \item First, the plate and the stiffener are meshed independently and the location of the stiffener control points is identified. Take Figure \ref{fig:VibMeshStiffPlate} for instance, where the control point A belongs to the plate element $k$. 
    \FloatBarrier \begin{figure}[htbp]
\centering
\includegraphics[scale=.5]{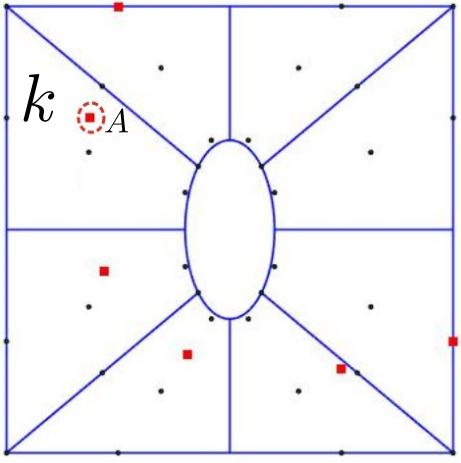}
\caption{Mesh plot of the curvilinearly stiffened plate in the physical space.}
\label{fig:VibMeshStiffPlate}
\end{figure} \FloatBarrier
    
    We can compute the natural coordinates ($\bar{\xi}, \bar{\eta}$)  of the stiffener control point in the plate element by solving Equation \ref{eq:vibnaturalcoordinates} since the stiffener and the plate control points are known.
    
    \begin{equation}
    \label{eq:vibnaturalcoordinates}
    {\mathbf r}_\mathbf s\mid_{A}=\sum_{j=1}^{4}R_j{(\bar{\xi}_{A},\bar{\eta}_{A})}{\mathbf x}_{\mathbf pj}\mid_{k}.
\end{equation}

\item Since the natural coordinates are not obtained, the displacement of the stiffener is:

    \begin{equation}
    \label{eq:45}
    {\mathbf u}_\mathbf s\mid_{A}=\sum_{j=1}^{4}R_j{(\bar{\xi}_{A},\bar{\eta}_{A})}{\mathbf u}_{\mathbf pj}\mid_{k}.
\end{equation}

\item The displacements of any points in the stiffener element can be expressed as:
    \begin{equation}
    \label{eq:46}
    {\mathbf u}_\mathbf {sg}=\sum_{j=A}R_j{(\varsigma)}{\mathbf u}_{\mathbf sj}.
\end{equation}

\item Then, substituting Equation \ref{eq:45} into Equation \ref{eq:46} results in the following equation:
\begin{equation}
    \mathbf u_\mathbf {sg}=\sum_{j=A}R_j{(\varsigma)}\sum_{i=1}^{4}R_i(\bar{\xi}_{j},\bar{\eta}_{j}){\mathbf u}_{\mathbf p i}\mid_{k}
\end{equation}

\item Rewriting the displacement approximations in a matrix form,
\begin{equation}
    \mathbf u_\mathbf {sg}=\mathbf N_{ps}{\mathbf u}_{\mathbf p}\mid_{k}.
    \end{equation}

\item Now the displacement $\mathbf{u}_{sg}$ are described in the global coordinate system. Hence the stiffener displacements should be transformed to the local curvilinear coordinate system. The transformation matrix is:
\begin{equation}
    \mathbf T={\left(\begin{array}{ccccc}\cos\alpha&\sin\alpha&0&0&0\\-\sin\alpha&\cos\alpha&0&0&0\\0&0&1&0&0\\0&0&0&\cos\alpha&\sin\alpha\\0&0&0&-\sin\alpha&\cos\alpha\end{array}\right)},
\end{equation}
where $\alpha$ is the angle between the stiffener tangential direction $t$-axis and $x$-axis of the global coordinate system, whose value is calculated at each integration point.
\begin{equation}
\alpha=tan^{-1}\Big(\frac{y'}{x'}\Big)
\end{equation}

The prime ' denotes the first derivative with respect to the stiffener arc length, $d()/d\Gamma$.
\end{itemize}

Hence, the stiffener strain energy $U_s$ and the kinetic energy $T_s$ can be rewritten as,

\begin{equation}
    U_s=\frac12\int\mathbf u_\mathbf p^\mathrm T\mathbf N_\mathbf{ps}^\mathrm T\mathbf T^\mathrm T\mathbf B_\mathbf s^{\mathbf L\mathrm T}{\mathbf D}_\mathbf s\mathbf B_\mathbf s^\mathbf L\mathbf T{\mathbf N}_\mathbf{ps}{\mathbf u}_\mathbf p\mathrm d\Gamma,
\end{equation}

\begin{equation}
    T_s=\frac12\int\dot{\mathbf u}_\mathbf p^\mathrm T\mathbf N_\mathbf{ps}^\mathrm T\mathbf T^\mathrm T{\mathbf m}_\mathbf s\mathbf T{\mathbf N}_\mathbf{ps}{\dot{\mathbf u}}_\mathbf p\mathrm d\Gamma,
\end{equation}

Hamilton's principle can now be used in deriving the weak form of the governing equation.

\begin{equation}
    \int_{t_1}^{t_2}{\left[{\left(\delta U_p+\delta U_s\right)}-{\left(\delta T_p+\delta T_s\right)}\right]}\mathrm dt=0,
\end{equation}
where $t_1$ and $t_2$ are the initial and final times, respectively.
The principle expression is:

\begin{equation}
\int_{t_1}^{t_2}{\left[\begin{array}{c}\underset\Omega{\hspace{0.28em}\int\int\hspace{0.28em}}{\delta\mathbf u_\mathbf p^\mathrm T{(\mathbf B}_\mathbf p^{\mathbf L\mathrm T}{\mathbf D}_\mathbf p\mathbf B_\mathbf p^\mathbf L){\mathbf u}_\mathbf p}\mathrm d\Omega+\int{\delta\mathbf u_\mathbf p^\mathrm T{(\mathbf N}_\mathbf{ps}^\mathrm T\mathbf T^\mathrm T\mathbf B_\mathbf s^{\mathbf L\mathrm T}{\mathbf D}_\mathbf s\mathbf B_\mathbf s^\mathbf L\mathbf T{\mathbf N}_\mathbf{ps}){\mathbf u}_\mathbf p\mathrm d\Gamma}\\\hspace{0.28em}\hspace{0.28em}\hspace{0.28em}\hspace{0.28em}\hspace{0.28em}\hspace{0.28em}\hspace{0.28em}\hspace{0.28em}\hspace{0.28em}\hspace{0.28em}\hspace{0.28em}\hspace{0.28em}-\underset\Omega{\hspace{0.28em}\int\int\hspace{0.28em}}\delta\mathbf u_\mathbf p^\mathrm T{\mathbf m}_\mathbf p{\ddot{\mathbf u}}_\mathbf p\mathrm d\Omega-\int{\delta\mathbf u_\mathbf p^\mathrm T{(\mathbf N}_\mathbf{ps}^\mathrm T\mathbf T^\mathrm T{\mathbf m}_\mathbf s\mathbf T{\mathbf N}_\mathbf{ps}){\ddot{\mathbf u}}_\mathbf p\mathrm d\Gamma}\end{array}\right]}\mathrm dt=0.
\end{equation}

To exhibit and substantiate the technique, free vibration analysis was conducted on multiple plates featuring diverse curvilinear stiffeners. The obtained outcomes were juxtaposed with established literature in certain instances, and in others, they were cross-referenced using widely recognized software such as ABAQUS. Through parametric investigations, the impact of both the cutout contour and the configuration and placement of the curvilinear stiffeners on the results was elucidated.

    \subsection{Modeling complicated cutouts using a single NURBS patch} \label{ss:singlepatch}
\begin{figure}[htbp]
\centering
  \includegraphics[width=0.6\textwidth]{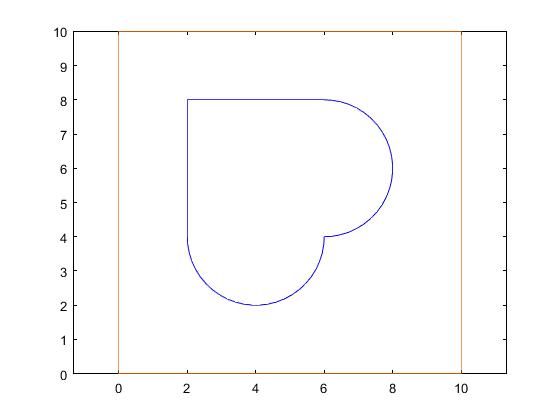}
\caption{ Inner and the outer curve to generate a plate with a complicated cutout}
\label{fig:InnerOuterCurve}
\end{figure}

\begin{figure}[htbp]
\centering
  \subfigure[Mesh Plot]{\includegraphics[width=3in]{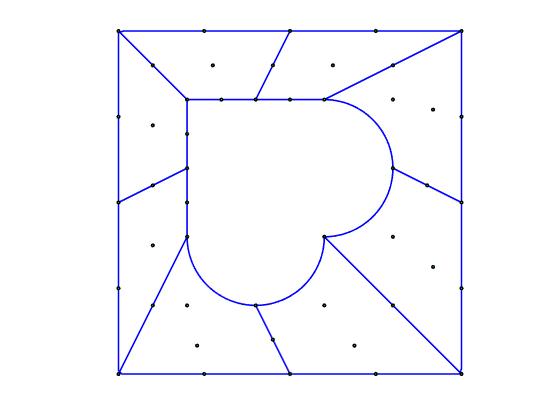}}
  \subfigure[Connectivity Plot]{\includegraphics[width=3in]{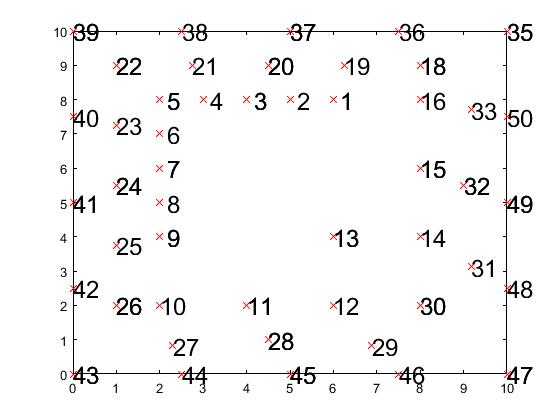}}
  
\caption{ Mesh and connectivity plot of a plate with a complicated cutout}
\label{fig:newconnplot}
\end{figure}

     In this case, first the inner and the outer profile of such a panel is constructed using 1D NURBS curves as shown in Figure \ref{fig:InnerOuterCurve}. The control points and weights used to construct these curves are detailed in Table \ref{tab:innercurve} and Table \ref{tab:outercurve}. Figure \ref{fig:newconnplot} and Table \ref{tab:newconn} describe the mesh and the connectivity matrix for the coarsest mesh respectively.
     
     \FloatBarrier \begin{table}[htbp]
\caption{Control points and weights of the inner curve}
\label{tab:innercurve}
\begin{tabular}{|c|c|c|c|c|c|c|c|c|c|c|c|c|c|c|c|c|c|c|}
\hline
\multirow{3}{*}{$P_{ij}$} & $x$ & 6 & 5 & 4 & 3 & 2 & 2 & 2 & 2 & 2 & 1.414 & 4 & 4.243 & 6 & 5.657 & 8 & 5.657 & 6 \\ \cline{2-19} 
 & $y$ & 8 & 8 & 8 & 8 & 8 & 7 & 6 & 5 & 4 & 1.414 & 2 & 1.414 & 4 & 2.828 & 6 & 5.657 & 8 \\ \cline{2-19} 
 & $z$ & 0 & 0 & 0 & 0 & 0 & 0 & 0 & 0 & 0 & 0 & 0 & 0 & 0 & 0 & 0 & 0 & 0 \\ \hline
\multicolumn{2}{|c|}{$W$} & 1 & 1 & 1 & 1 & 1 & 1 & 1 & 1 & 1 & 0.707 & 1 & 0.707 & 1 & 0.707 & 1 & 0.707 & 1 \\ \hline
\end{tabular}
\end{table} \FloatBarrier

\FloatBarrier \begin{table}[htbp]
\centering
\caption{Control points and weights of the outer curve}
\label{tab:outercurve}
\begin{tabular}{|c|c|c|c|c|c|c|c|c|c|c|}
\hline
\multirow{3}{*}{$P_{ij}$} & $x$ & 10 & 5 & 0 & 0 & 0 & 5 & 10 & 10 & 10 \\ \cline{2-11} 
 & $y$ & 10 & 10 & 10 & 5 & 0 & 0 & 0 & 5 & 10 \\ \cline{2-11} 
 & $z$ & 0 & 0 & 0 & 0 & 0 & 0 & 0 & 0 & 0 \\ \hline
\multicolumn{2}{|c|}{$W$} & 1 & 1 & 1 & 1 & 1 & 1 & 1 & 1 & 1 \\ \hline
\end{tabular}
\end{table} \FloatBarrier
\FloatBarrier  \FloatBarrier

\FloatBarrier     \FloatBarrier

\FloatBarrier \begin{table}[htbp]
\caption{Connectivity matrix of a plate with a complicated cutout}
\centering
\begin{tabular}{|c|c|c|c|c|c|c|c|c|c|}
\hline
\textbf{Element} & \multicolumn{9}{c|}{\textbf{Node}} \\ \hline
\textbf{} & \textbf{1} & \textbf{2} & \textbf{3} & \textbf{4} & \textbf{5} & \textbf{6} & \textbf{7} & \textbf{8} & \textbf{9} \\ \hline
\textbf{1} & 1 & 2 & 3 & 18 & 19 & 20 & 35 & 36 & 37 \\ \hline
\textbf{2} & 3 & 4 & 5 & 20 & 21 & 22 & 37 & 38 & 39 \\ \hline
\textbf{3} & 5 & 6 & 7 & 22 & 23 & 24 & 39 & 40 & 41 \\ \hline
\textbf{4} & 7 & 8 & 9 & 24 & 25 & 26 & 41 & 42 & 43 \\ \hline
\textbf{5} & 9 & 10 & 11 & 26 & 27 & 28 & 43 & 44 & 45 \\ \hline
\textbf{6} & 11 & 12 & 13 & 28 & 29 & 30 & 45 & 46 & 47 \\ \hline
\textbf{7} & 13 & 14 & 15 & 30 & 31 & 32 & 47 & 48 & 49 \\ \hline
\textbf{8} & 15 & 16 & 1 & 32 & 33 & 18 & 49 & 50 & 35 \\ \hline
\end{tabular}
\label{tab:newconn}
\end{table} \FloatBarrier

    \subsection{A Three point parametrization of the curvilinear stiffener} \label{ss:ls_multipatchsec}
Previous researchers \cite{Tamijani2010}, \cite{Zhao2016} and \cite{Shi2015} involved curvilinear stiffeners with a parabolic profile. Using the parametric equation of the parabola and the start and the end point coordinates, one can obtain the three control points which could define the Bezier curve \cite{Piegl1996}. Since any parabolic stiffener can be represented using three control points, the coordinates of the end control points $\Delta\epsilon$ and Point 2 (where  $\delta_{dist}$  refers to the point [$\delta_{dist}$ ,$\delta_{dist}$]) were used to represent the curvilinear stiffener.
\FloatBarrier \begin{figure}[htbp]
\centering
  \includegraphics[width=.5\textwidth]{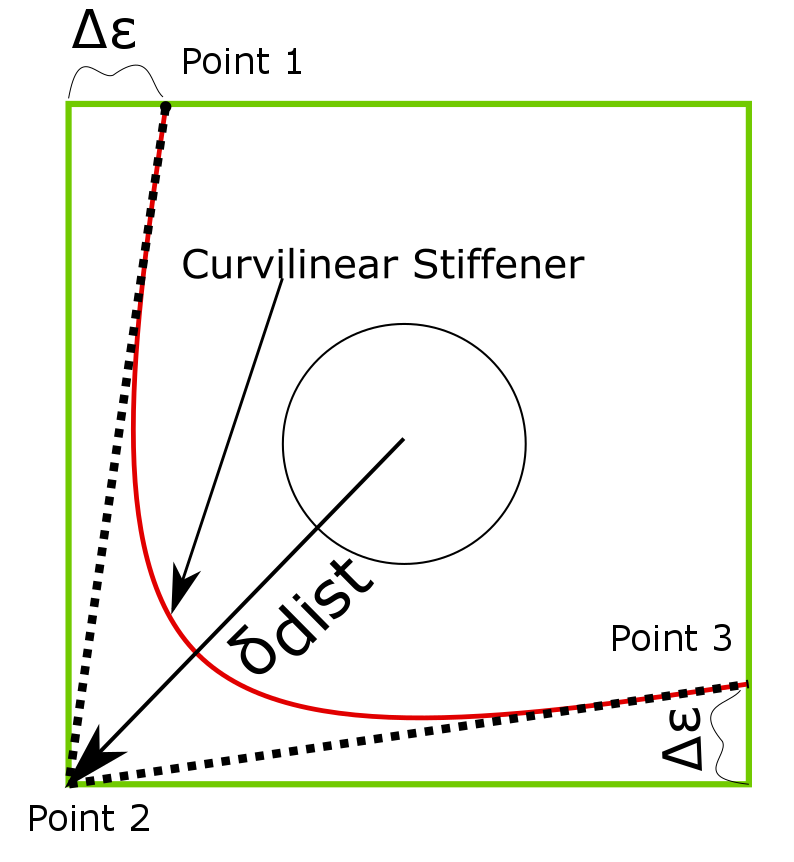}
\caption{Parametrization of curvilinear stiffener}
\label{fig:ls_Parametrization}
\end{figure} \FloatBarrier

The FSDT which relies on $C^{0}$  continuity of the basis
functions is employed in the patch interior unlike Kirchhoff-Love shell theory that relies on higher-order continuity of the basis
functions as mentioned in \cite{Bazilevs2010b}.

\section{Results and Discussion} \label{se:one_section}
To demonstrate and  validate the method, free vibration analysis of several plates with different curvilinear stiffeners were carried out. Results are compared with existing literature for some cases while for some others, using commercially popular software like ABAQUS. Parametric studies were performed which show the influence of both the cutout profile and the shape and position of the curvilinear stiffeners in the results.

\subsection{Isotropic plate with a circular hole at the center}
In order to illustrate the performance of IGA code in modeling holes, an isotropic plate with a circular hole at the center was considered. The plate is assumed to have side length of 10 m and a thickness of 0.1 m. The hole has a radius of 1 m,.  
\newline 
\begin{equation}
\mathit{SSSS}:\left\{\begin{array}{c}u_0=v_0=w_0=\phi_y=\phi_z=0\hspace{1em}\text{on}\hspace{0.35em}x\hspace{0.35em}=\hspace{0.35em}0\text{,}\hspace{0.35em}a\\u_0=v_0=w_0=\phi_x=\phi_z=0\hspace{1em}\text{on}\hspace{0.35em}y\hspace{0.35em}=\hspace{0.35em}0\text{,}\hspace{0.35em}b\end{array}\right.
\end{equation}
The material properties are :
\begin{equation}
\begin{array}{c}E\hspace{0.35em} =\hspace{0.35em} 208\hspace{0.35em} GPa\text{,}\hspace{0.35em}\nu =\hspace{0.35em} 0.3\text{,}\hspace{0.35em}\rho =\hspace{0.35em} 8,000kg/m^2\end{array}\end{equation}

A normalized frequency parameter is defined by $\widetilde\omega={\left[\rho h\omega^2a^4/D{\left(1-v^2\right)}\right]}^{1/4}$  with $D=Eh^3/12{\left(1-v^2\right)}$. 

\begin{table}[htbp]
\centering
\caption{Normalized frequencies of clamped square plate with a circular hole.}
\label{tab:circcutout}
\begin{tabular}{|c|c|c|c|}
\hline
\textbf{Mode} & \textbf{Present} & \cite{Huang1999} & \cite{Liu2008} \\ \hline
\textbf{1} & 6.183 & 6.24 & 6.149 \\ \hline
\textbf{2} & 8.657 & 8.457 & 8.577 \\ \hline
\textbf{3} & 8.657 & 8.462 & 8.634 \\ \hline
\textbf{4} & 10.513 & 10.233 & 10.422 \\ \hline
\textbf{5} & 11.559 & 11.719 & 11.414 \\ \hline
\textbf{6} & 12.052 & 12.299 & 11.838 \\ \hline
\textbf{7} & 13.022 & 13.037 & 12.829 \\ \hline
\textbf{8} & 13.022 & 13.041 & 12.842 \\ \hline
\end{tabular}
\end{table}

The first eight frequencies obtained by the current method are shown in Table \ref{tab:circcutout}. These results are compared against reference solutions in \cite{Huang1999}, \cite{Liu2008}. The results are seen to be in good agreement with the reference solution.
\subsection{Isotropic plate with a complicated cutout}
The present method is applied to analyze an isotropic plate with a heart shaped cutout (Figure \ref{fig:InnerOuterCurve}). The thickness of the plate is $h = 0.05 $m and the material parameters are
\begin{equation}
\begin{array}{c}E\hspace{0.35em} =\hspace{0.35em} 200\hspace{0.35em} GPa\text{,}\hspace{0.35em}\nu =\hspace{0.35em} 0.3\text{,}\hspace{0.35em}\rho =\hspace{0.35em} 8000 \hspace{0.35em} kg/m^3\end{array}\end{equation}
Shojaee et al. \cite{shojaee2012free} analyzed the same geometry using IGA and classical laminate theory with multiple patches. The bending strip method was used to achieve compatibility between adjacent patches. In this work, the first order shear deformation theory was used and the plate was modeled with a single NURBS patch thus eliminating the use of multiple NURBS patches.
\begin{table}[htbp]
\centering
\caption{Normalized natural frequencies of a simply supported plate with a heart shaped complicated cutout.}
\label{tab:SSSS}
\begin{tabular}{|c|c|c|c|c|c|c|}
\hline
\textbf{Mode} & \textbf{MultiPatch} & \textbf{MKI} & \textbf{EFG} & \textbf{RIPM} & \textbf{ABAQUS} & \textbf{Present} \\ \hline
1 & 5.193 & 5.39 & 5.453 & 4.919 & 4.948 & 4.945 \\ \hline
2 & 6.579 & 7.502 & 8.069 & 6.398 & 6.426 & 6.441 \\ \hline
3 & 6.597 & 8.347 & 9.554 & 6.775 & 6.796 & 6.804 \\ \hline
4 & 7.819 & 10.636 & 10.099 & 8.613 & 8.616 & 8.585 \\ \hline
5 & 8.812 & 11.048 & 11.328 & 9.016 & 9.020 & 9.031 \\ \hline
6 & 9.42 & 12.894 & 12.765 & 10.738 & 10.720 & 10.670 \\ \hline
7 & 10.742 & 13.71 & 13.685 & 10.93 & 10.966 & 10.908 \\ \hline
8 & 10.776 & 14.062 & 14.305 & 11.601 & 11.686 & 11.661 \\ \hline
9 & 11.919 & 16.649 & 15.721 & 12.903 & 12.901 & 12.833 \\ \hline
10 & 13.2 & 17.364 & 17.079 & 13.283 & 13.238 & 13.185 \\ \hline
\end{tabular}
\end{table}
\begin{figure}[htbp]
\centering
\begin{tikzpicture}[scale=1.2]
\tikzstyle{every node}=[font=\small]
\begin{axis}[xmin=1, xmax=10,
ymin=0, ymax=30,
xlabel={Modes},
ylabel={Nondimensional Frequency},legend style={nodes={scale=0.5, transform shape}}]
\addlegendimage{empty legend}
\addplot coordinates {
(	1	,	5.193	)
(	2	,	6.579	)
(	3	,	6.597	)
(	4	,	7.819	)
(	5	,	8.812	)
(	6	,	9.42	)
(	7	,	10.742	)
(	8	,	10.776	)
(	9	,	11.919	)
(	10	,	13.2	)

};
\addplot coordinates{
(	1	,	5.39	)
(	2	,	7.502	)
(	3	,	8.347	)
(	4	,	10.636	)
(	5	,	11.048	)
(	6	,	12.894	)
(	7	,	13.71	)
(	8	,	14.062	)
(	9	,	16.649	)
(	10	,	17.364	)
};
\addplot coordinates{

(	1	,	5.453	)
(	2	,	8.069	)
(	3	,	9.554	)
(	4	,	10.099	)
(	5	,	11.328	)
(	6	,	12.765	)
(	7	,	13.685	)
(	8	,	14.305	)
(	9	,	15.721	)
(	10	,	17.079	)
};
\addplot coordinates{

(	1	,	4.919	)
(	2	,	6.398	)
(	3	,	6.775	)
(	4	,	8.613	)
(	5	,	9.016	)
(	6	,	10.738	)
(	7	,	10.93	)
(	8	,	11.601	)
(	9	,	12.903	)
(	10	,	13.283	)
};
\addplot coordinates{
(	1	,	4.948	)
(	2	,	6.426	)
(	3	,	6.796	)
(	4	,	8.616	)
(	5	,	9.020	)
(	6	,	10.720	)
(	7	,	10.966	)
(	8	,	11.686	)
(	9	,	12.901	)
(	10	,	13.238	)
};
\addplot coordinates{

(	1	,	4.945	)
(	2	,	6.441	)
(	3	,	6.804	)
(	4	,	8.585	)
(	5	,	9.031	)
(	6	,	10.670	)
(	7	,	10.908	)
(	8	,	11.661	)
(	9	,	12.833	)
(	10	,	13.185	)

};

   \addlegendentry{\textbf{Methods}}
   \addlegendentry{Multipatch}
   \addlegendentry{MKI}
   \addlegendentry{EFG}
   \addlegendentry{RPIM}
   \addlegendentry{ABAQUS}
   \addlegendentry{Present}
   
\end{axis}
\end{tikzpicture}
\caption{Normalized natural frequencies of a simply supported plate with a heart shaped complicated cutout.} \label{fig:pointtwofive}
\end{figure}
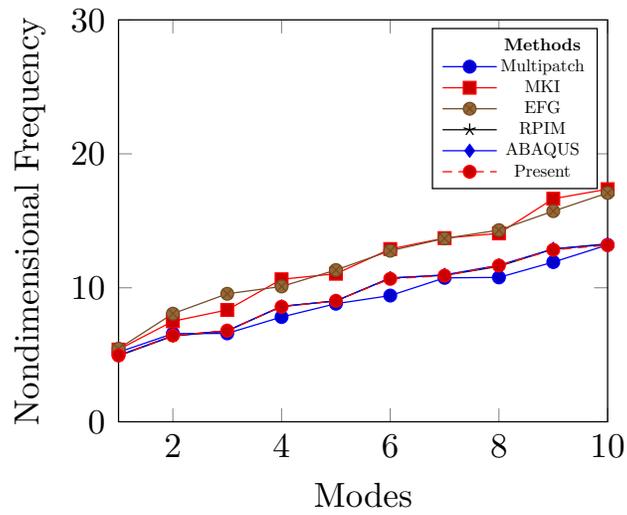

\begin{figure}[htbp]
\centering
  \includegraphics[width=0.7\textwidth]{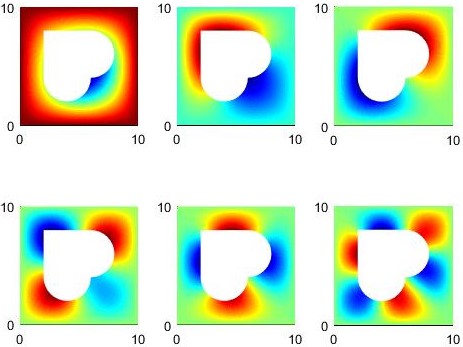}
\caption{ First six modes (IGA) of a simply supported square plate with a complicated cutout}
\label{fig:vibration}
\end{figure}

\begin{figure}[htbp]
\centering
  \includegraphics[width=0.6\textwidth]{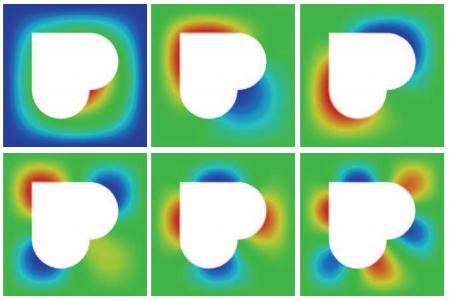}
\caption{First six modes \cite{shojaee2012free} of a simply supported plate with a heart shaped complicated cutout}
\label{fig:VibrationReference}
\end{figure}
The normalized frequencies  obtained using the present method are compared against some results from existing literature \cite{shojaee2012free}, MKI method \cite{Bui2011}, radial point interpolation method \cite{cui2011thin} and Element Free Galerkin method \cite{liu2001mesh}. The results can be seen in good agreement with the reference solutions for the simply supported boundary condition. As opposed to \cite{shojaee2012free}, the current method obviates the necessity to use the bending strip methods for patch coupling. The overall computational cost required is hence reduced significantly.
\subsection{Vibration of a Composite plate with a heart shaped cutout}
A three-layered symmetric crossply laminated composite plate is considered Figure \ref{fig:InnerOuterCurve}. The ratio of elastic constants are as follows:
\begin{equation}
\begin{array}{c}E_L/E_T=2.45\text{,}\hspace{0.3em}G_\mathit{LT}/E_T=0.48\text{,}\hspace{0.3em}G_\mathit{TT}/E_T=0.2\text{,}\hspace{0.3em}\nu_\mathit{LT}=0.23\end{array}\end{equation}

mass density = 8,000 $kg/m^3$ and thickness $h$ = $0.06 m$. The frequency is normalized using $\widetilde\omega={\left[\rho h\omega^2a^4/D_{0.1}\right]}^{1/2}$ , with $a$=$10 m$ and $D_{0.1}=E_1h^3/12{\left(1-v_{12}v_{21}\right)}$. The results were obtained for simply supported (SSSS) boundary condition.
\begin{equation}
\mathit{SSSS}:\left\{\begin{array}{c}u_0=v_0=w_0=\phi_y=\phi_z=0\hspace{1em}\text{on}\hspace{0.35em}x\hspace{0.35em}=\hspace{0.35em}0\text{,}\hspace{0.35em}a\\u_0=v_0=w_0=\phi_x=\phi_z=0\hspace{1em}\text{on}\hspace{0.35em}y\hspace{0.35em}=\hspace{0.35em}0\text{,}\hspace{0.35em}b\end{array}\right.
\end{equation} The results were obtained for simply supported (SSSS) boundary conditions. The first six normalized frequencies of this model for various angle ply fiber orientations are presented in Table \ref{tab:ssssvibtab}. The present solutions are compared with the results obtained by IGA and Kirchhoff theory with eight patches \cite{shojaee2012free}, EFG and MKI methods \cite{Bui2011}. It can be observed that the present method delivers very good results which are in agreement with results from published literature for considered angle ply orientations. 
 \FloatBarrier \begin{table}[htbp]
 \caption{Normalized natural frequencies of a simply supported laminated  plate with a heart shaped cutout for various angle ply orientations.}
 \label{tab:ssssvibtab}
\begin{tabular}{|c|c|c|c|c|c|c|c|}
\hline
Angle ply & Method & \multicolumn{6}{c|}{Mode} \\ \hline
 &  & 1 & 2 & 3 & 4 & 5 & 6 \\ \hline
\multirow{4}{*}{(15/−15°/15°)} & Present & 18.91 & 31.83 & 36.09 & 57.00 & 62.73 & 83.93 \\ \cline{2-8} 
 & 8-patch IGA\cite{shojaee2012free} & 18.91 & 32.05 & 36.00 & 56.35 & 63.37 & 83.63 \\ \cline{2-8} 
 & EFG\cite{Bui2011}& 19.18 & 32.45 & 37.24 & 58.72 & 63.99 & 86.50 \\ \cline{2-8} 
 & MKI\cite{Bui2011}& 18.32 & 31.47 & 37.62 & 63.08 & 66.54 & 86.49 \\ \hline
\multirow{4}{*}{(30/−30°/30°)} & Present & 20.40 & 33.66 & 37.23 & 59.20 & 65.03 & 87.92 \\ \cline{2-8} 
 & 8-patch IGA\cite{shojaee2012free} & 20.32 & 33.93 & 37.07 & 58.48 & 65.90 & 87.97 \\ \cline{2-8} 
 & EFG\cite{Bui2011}& 20.93 & 34.92 & 39.10 & 62.22 & 67.05 & 92.72 \\ \cline{2-8} 
 & MKI\cite{Bui2011}& 20.31 & 33.99 & 39.90 & 58.11 & 69.70 & 92.10 \\ \hline
\multirow{4}{*}{(45°/−45°/45°)} & Present & 21.10 & 34.45 & 37.91 & 60.29 & 66.22 & 90.68 \\ \cline{2-8} 
 & 8-patch IGA\cite{shojaee2012free} & 20.98 & 34.85 & 37.56 & 59.33 & 67.52 & 91.22 \\ \cline{2-8} 
 & EFG\cite{Bui2011}& 21.74 & 36.08 & 39.98 & 63.90 & 68.53 & 96.77 \\ \cline{2-8} 
 & MKI\cite{Bui2011} & 20.99 & 34.90 & 39.27 & 63.38 & 69.02 & 96.59 \\ \hline
\end{tabular}
\end{table}  \FloatBarrier

\subsection{Curvilinearly stiffened composite panels with central circular cutout}
\label{ss:vibcurvmat}
\begin{figure}[htbp]
\begin{longtable}{ >{\centering\arraybackslash} m{0.25cm} >{\centering\arraybackslash} m{4cm} >{\centering\arraybackslash} m{0.7cm} >{\centering\arraybackslash} m{4cm} >{\centering\arraybackslash} m{0.7cm}}
\hline 
 & \multicolumn{2}{c}{Present}    & \multicolumn{2}{c}{ABAQUS}    \\
\hline \\
Mode & & $\widetilde\omega$ & & $\widetilde\omega$\\ \hline
1  & \includegraphics[scale=.32]{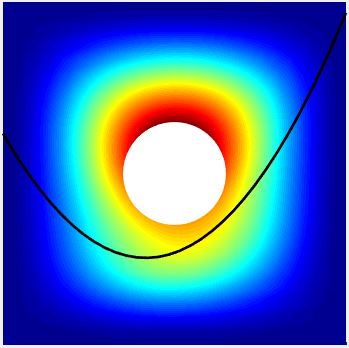} & 26.839
 &  \includegraphics[scale=.22]{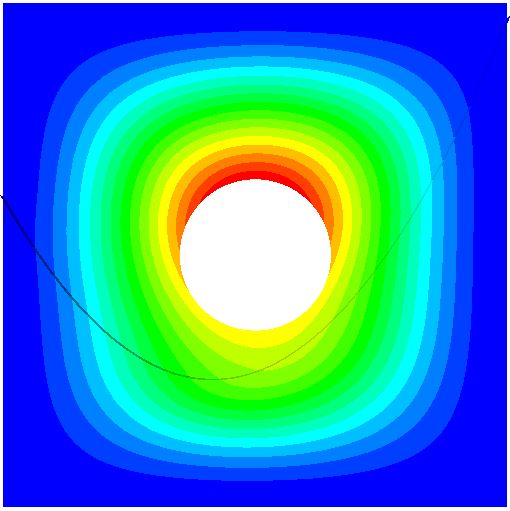} & 26.215\\
 2 & \includegraphics[scale=.32]{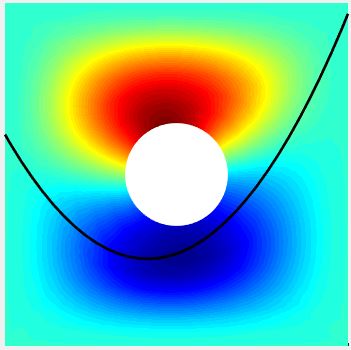} & 35.683
 &  \includegraphics[scale=.22]{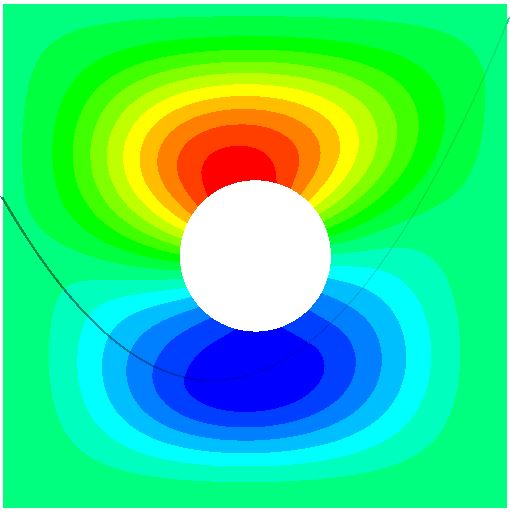} & 34.956
\\
 3 & \includegraphics[scale=.32]{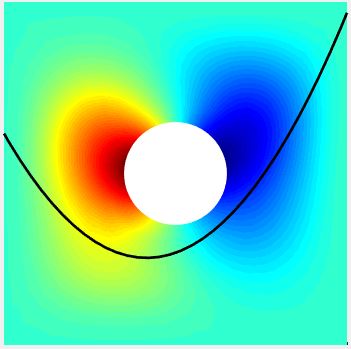} & 52.254
 &  \includegraphics[scale=.22]{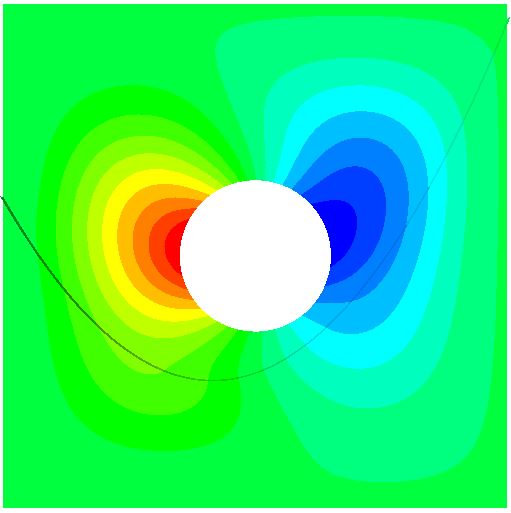} & 50.876
\\
 4 & \includegraphics[scale=.32]{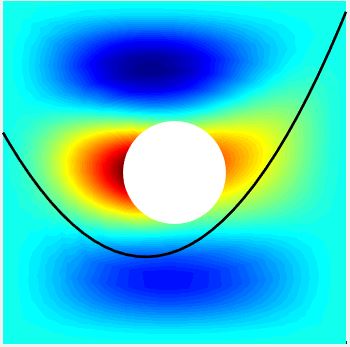} & 65.551
 &  \includegraphics[scale=.22]{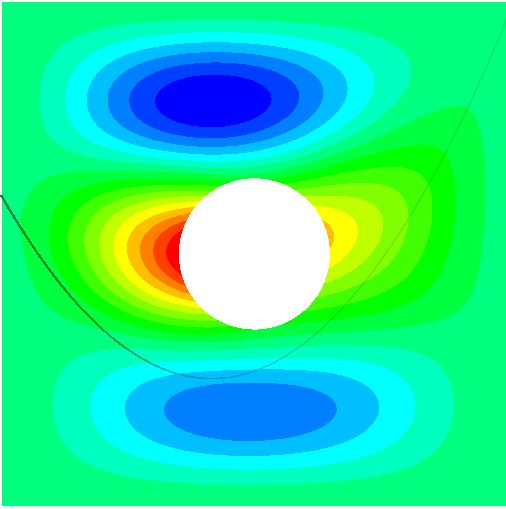} & 64.421
\\
 5 & \includegraphics[scale=.32]{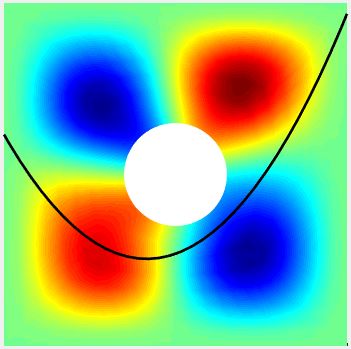} & 69.485
 &  \includegraphics[scale=.22]{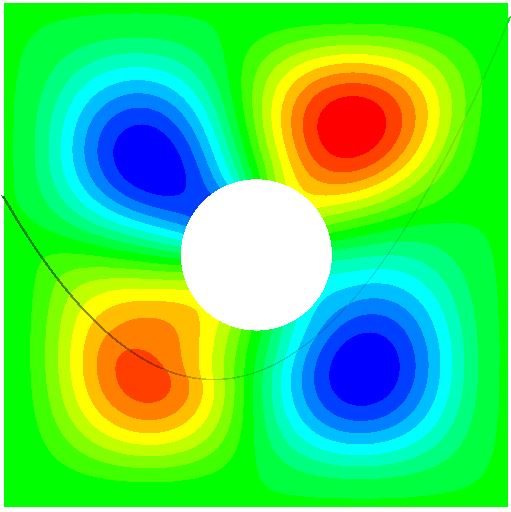} & 67.907
\\
\endfirsthead
\endhead

\end{longtable}
\caption{The first five eigenmode shape plots for a plate with curvilinear stiffener using a IGA and ABAQUS, a commercial
available software.}
\label{fig:vibmp_modeplot}
\end{figure}

In this section, the influence of central cutouts on clamped stiffened composite panels is studied. 
\begin{equation}
Clamped : \hspace{0.35em}
u_0=0,v_0=0,\psi_x=0,\psi_y=0, w_0=0
\end{equation}
on all four edges.
The stiffener is assumed to be fabricated from isotropic material with Young's modulus = $E_{T}$, Poisson's ratio = $\nu_{LT}$ and coefficient of thermal expansion = $\alpha_{0}$. The stiffness ratio $\gamma=\mathit{EI}/\mathit{bD}$  and the area ratio $\delta=A_s/{\mathit{bt}}_p$ are 5 and 0.1 respectively, unless specified otherwise. The ratio of elastic constants are as follows:
\begin{equation}
\begin{array}{c}E_L/E_T=15\text{,}\hspace{0.3em}G_\mathit{LT}/E_T=0.5\text{,}\hspace{0.3em}G_\mathit{TT}/E_T=0.3356\text{,}\hspace{0.3em}\nu_\mathit{LT}=0.3\end{array}\end{equation}

mass density = 8,000 $kg/m^3$ and thickness $h$ = $0.01$ $m$. The frequency is normalized using 
\begin{equation}
\widetilde\omega={\left[\rho h\omega^2a^4/D_{0.1}\right]}^{1/2}
\end{equation} with $a$=$1$ $m$ and $D_{0.1}=E_1h^3/12{\left(1-v_{12}v_{21}\right)}$. To check the accuracy of the displacement compatibility algorithm, results for clamped curvilinearly stiffened composite plate are compared with those obtained
using ABAQUS. The curvilinear stiffener configuration was adopted from \cite{tamijani2010buckling} and scaled proportionally to match the dimensions of the plate. From the ABAQUS library, the plate was meshed using 2040 S8R elements and curvilinear stiffener was meshed using 192 elements S8R elements (48 divisions along the arc length and 4 divisions along the depth direction). The IGA plate model was meshed using 1024 elements and the curvilinear stiffener using 32 elements. The normalized fundamental frequency was observed to converge for these mesh configurations. The first five eigenmodes were extracted for this case. The eigenmode plots obtained using the IGA approach
are similar to the ones obtained using ABAQUS as can be seen from Figure \ref{fig:vibmp_modeplot}.

\pagebreak

\FloatBarrier  \FloatBarrier

\subsection{Curvilinearly stiffened composite panels with an elliptical central cutout}
In this section, results of curvilinearly stiffened composite panels with elliptical cutouts are presented.
The ratio of elastic constants are as follows:
\begin{equation}
\begin{array}{c}E_L/E_T=15\text{,}\hspace{0.3em}G_\mathit{LT}/E_T=0.5\text{,}\hspace{0.3em}G_\mathit{TT}/E_T=0.3356\text{,}\hspace{0.3em}\nu_\mathit{LT}=0.3\end{array}\end{equation}

mass density = 8,000 $kg/m^3$ and thickness $h$ = $0.01$ $m$. The frequency is normalized by $\widetilde\omega={\left[\rho h\omega^2a^4/D_{0.1}\right]}^{1/2}$ , with $a$=$1$ $m$ and $D_{0.1}=E_1h^3/12{\left(1-v_{12}v_{21}\right)}$. The results were obtained for clamped boundary condition.

An ellipse of semi-major axis = $0.2$ $m$ and semi-minor axis = $0.1$ $m$  is considered for all cases. The  normalized first ten frequencies of a stiffend four layer composite laminate plate with respect to ply orientations, stiffness ratios and different stiffener profiles (defined by $\Delta\epsilon$ Figure \ref{fig:ls_Parametrization}) are presented in Table \ref{tab:vib1} and Table \ref{tab:vib2} and their fifth mode shapes are shown in Figures \ref{tab:vib1}, \ref{tab:vib2}, \ref{tab:vib3}, \ref{tab:vib4}, \ref{tab:vib5} and \ref{tab:vib6}.

\FloatBarrier \begin{table}[htbp]
\caption {Normalized frequencies for different ply orientations for $\gamma=5$ and $\Delta\epsilon=0$} \label{tab:vib1} 
\centering
\begin{tabular}{|c|c|c|c|c|}
\hline
\textbf{Mode} & \textbf{\begin{tabular}[c]{@{}c@{}}Antisymmetric\\ Crossply\end{tabular}} & \textbf{\begin{tabular}[c]{@{}c@{}}Symmetric\\ Crossply\end{tabular}} & \textbf{\begin{tabular}[c]{@{}c@{}}Symmetric\\ Angleply\end{tabular}} & \textbf{\begin{tabular}[c]{@{}c@{}}Antisymmetric\\ Angleply\end{tabular}} \\ \hline
1 & 23.986 & 25.407 & 23.572 & 23.582 \\ \hline
2 & 38.356 & 34.523 & 37.198 & 38.044 \\ \hline
3 & 44.347 & 47.911 & 39.978 & 42.451 \\ \hline
4 & 61.730 & 60.599 & 60.905 & 64.111 \\ \hline
5 & 76.728 & 64.747 & 71.808 & 73.009 \\ \hline
6 & 84.472 & 77.037 & 80.460 & 76.987 \\ \hline
7 & 92.642 & 88.814 & 86.570 & 83.375 \\ \hline
8 & 97.597 & 100.505 & 88.596 & 100.486 \\ \hline
9 & 113.239 & 112.436 & 106.260 & 110.241 \\ \hline
10 & 131.673 & 117.184 & 114.723 & 115.131 \\ \hline
\end{tabular} 
\end{table} \FloatBarrier
\FloatBarrier \begin{table}[htbp]
\centering
\caption {Normalized frequencies for different ply orientations for $\gamma=5$ and $\Delta\epsilon=0.25$} \label{tab:vib2} 
\begin{tabular}{|c|c|c|c|c|}
\hline
\textbf{Mode} & \textbf{\begin{tabular}[c]{@{}c@{}}Antisymmetric\\ Crossply\end{tabular}} & \textbf{\begin{tabular}[c]{@{}c@{}}Symmetric\\ Crossply\end{tabular}} & \textbf{\begin{tabular}[c]{@{}c@{}}Symmetric\\ Angleply\end{tabular}} & \textbf{\begin{tabular}[c]{@{}c@{}}Antisymmetric\\ Angleply\end{tabular}} \\ \hline
1 & 25.733 & 26.909 & 24.646 & 25.235 \\ \hline
2 & 37.732 & 35.836 & 36.676 & 37.132 \\ \hline
3 & 48.315 & 49.624 & 43.724 & 47.161 \\ \hline
4 & 69.413 & 61.422 & 68.170 & 69.795 \\ \hline
5 & 74.287 & 73.238 & 70.369 & 70.890 \\ \hline
6 & 90.708 & 85.848 & 87.059 & 84.483 \\ \hline
7 & 97.896 & 94.369 & 92.601 & 91.590 \\ \hline
8 & 112.175 & 106.567 & 100.993 & 110.461 \\ \hline
9 & 116.899 & 120.171 & 109.681 & 112.065 \\ \hline
10 & 134.109 & 126.330 & 116.890 & 124.329 \\ \hline
\end{tabular}
\end{table} \FloatBarrier
\FloatBarrier \begin{table}[htbp]
\centering
\caption {Normalized frequencies for different ply orientations for  $\gamma=10$ and $\Delta\epsilon=0$} \label{tab:vib3} 
\begin{tabular}{|c|c|c|c|c|}
\hline
\textbf{Mode} & \textbf{\begin{tabular}[c]{@{}c@{}}Antisymmetric\\ Crossply\end{tabular}} & \textbf{\begin{tabular}[c]{@{}c@{}}Symmetric\\ Crossply\end{tabular}} & \textbf{\begin{tabular}[c]{@{}c@{}}Symmetric\\ Angleply\end{tabular}} & \textbf{\begin{tabular}[c]{@{}c@{}}Antisymmetric\\ Angleply\end{tabular}} \\ \hline
1 & 24.558 & 25.884 & 24.338 & 24.204 \\ \hline
2 & 39.728 & 35.893 & 38.942 & 39.532 \\ \hline
3 & 44.829 & 48.349 & 40.774 & 42.883 \\ \hline
4 & 63.772 & 62.179 & 62.164 & 65.975 \\ \hline
5 & 78.058 & 66.348 & 73.699 & 74.667 \\ \hline
6 & 86.129 & 79.246 & 81.491 & 77.525 \\ \hline
7 & 92.842 & 89.915 & 87.746 & 83.964 \\ \hline
8 & 99.651 & 100.505 & 88.801 & 101.741 \\ \hline
9 & 114.029 & 112.995 & 108.705 & 111.953 \\ \hline
10 & 134.182 & 119.439 & 115.836 & 118.263 \\ \hline
\end{tabular}
\end{table}
\FloatBarrier
\FloatBarrier
\begin{table}[htbp]
\centering
\caption {Normalized frequencies for different ply orientations for  $\gamma=10$ and $\Delta\epsilon=0.25$} \label{tab:vib4} 
\begin{tabular}{|c|c|c|c|c|}
\hline
\textbf{Mode} & \textbf{\begin{tabular}[c]{@{}c@{}}Antisymmetric\\ Crossply\end{tabular}} & \textbf{\begin{tabular}[c]{@{}c@{}}Symmetric\\ Crossply\end{tabular}} & \textbf{\begin{tabular}[c]{@{}c@{}}Symmetric\\ Angleply\end{tabular}} & \textbf{\begin{tabular}[c]{@{}c@{}}Antisymmetric\\ Angleply\end{tabular}} \\ \hline
1 & 27.711 & 28.829 & 26.494 & 27.386 \\ \hline
2 & 39.709 & 38.489 & 38.548 & 38.691 \\ \hline
3 & 52.994 & 51.992 & 48.197 & 52.150 \\ \hline
4 & 74.603 & 65.702 & 71.611 & 72.942 \\ \hline
5 & 77.687 & 79.184 & 76.178 & 74.211 \\ \hline
6 & 93.429 & 89.734 & 89.045 & 87.193 \\ \hline
7 & 102.543 & 98.080 & 95.125 & 99.094 \\ \hline
8 & 115.133 & 111.041 & 107.583 & 113.098 \\ \hline
9 & 126.002 & 121.882 & 113.833 & 114.182 \\ \hline
10 & 137.331 & 133.214 & 124.795 & 139.792 \\ \hline
\end{tabular}
\end{table}
\FloatBarrier

\FloatBarrier \begin{table}[htbp]
\centering
\caption {The fifth mode shape plots for different ply orientations for $\gamma=5$ and $\Delta\epsilon=0$} \label{tab:vib5} 
\centering
\begin{tabular}{|c|c|}
\hline
  \textbf{\begin{tabular}[c]{@{}c@{}}Antisymmetric Crossply\end{tabular}} & \textbf{\begin{tabular}[c]{@{}c@{}}Symmetric Crossply\end{tabular}} \\ \hline  \hline
		\includegraphics[scale=.35]{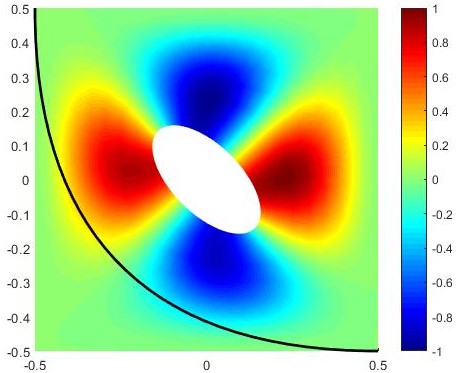}	&	\includegraphics[scale=.35]{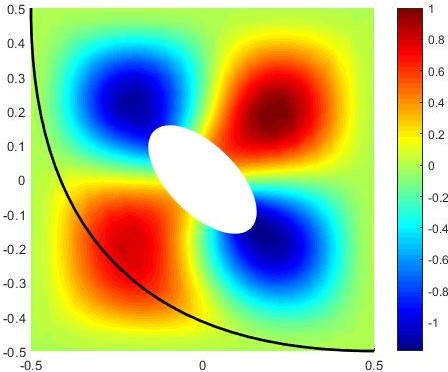}	\\ \hline
		\textbf{\begin{tabular}[c]{@{}c@{}}Symmetric Angleply\end{tabular}} & \textbf{\begin{tabular}[c]{@{}c@{}}Antisymmetric Angleply\end{tabular}} \\ \hline
		\includegraphics[scale=.35]{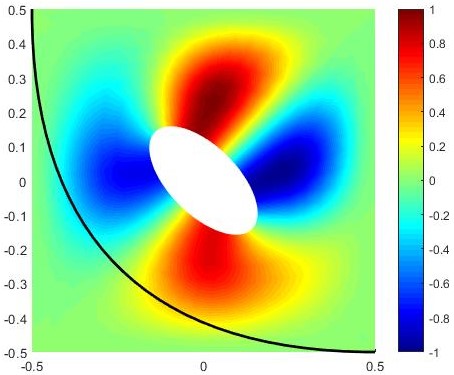}	&	\includegraphics[scale=.35]{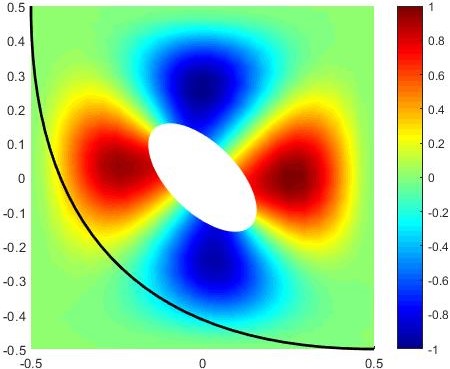}	\\	\hline
\end{tabular}
\end{table} \FloatBarrier
\FloatBarrier \begin{table}[htbp]
\centering
\caption {The fifth mode shape plots for different ply orientations for $\gamma=5$ and $\Delta\epsilon=0.25$} \label{tab:vib7} 
\centering
\begin{tabular}{|c|c|}
\hline
  \textbf{\begin{tabular}[c]{@{}c@{}}Antisymmetric Crossply\end{tabular}} & \textbf{\begin{tabular}[c]{@{}c@{}}Symmetric Crossply\end{tabular}} \\ \hline  \hline
		\includegraphics[scale=.35]{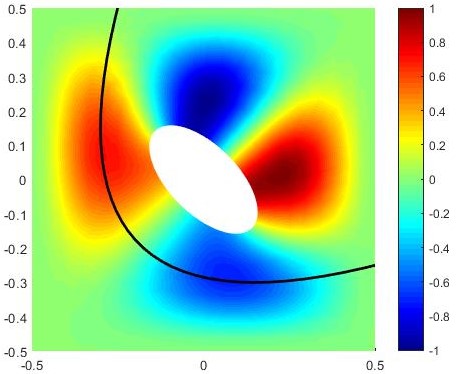}	&	\includegraphics[scale=.35]{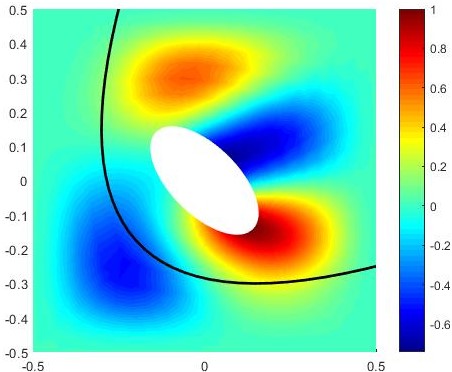}	\\ \hline
		\textbf{\begin{tabular}[c]{@{}c@{}}Symmetric Angleply\end{tabular}} & \textbf{\begin{tabular}[c]{@{}c@{}}Antisymmetric Angleply\end{tabular}} \\ \hline
		\includegraphics[scale=.35]{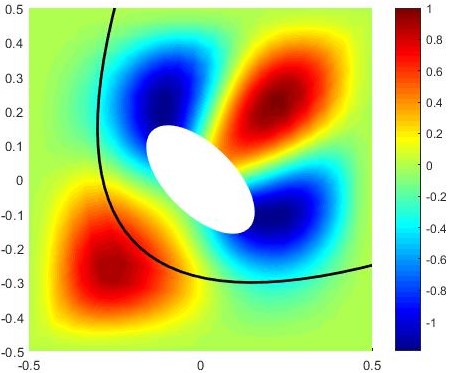}	&	\includegraphics[scale=.35]{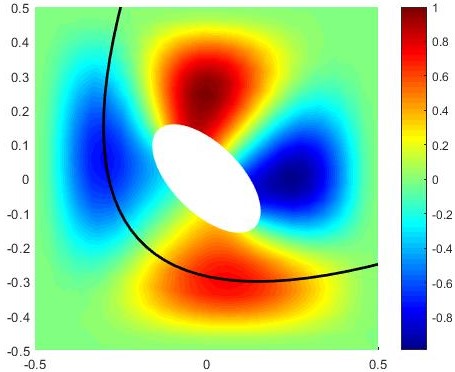}	\\	\hline
\end{tabular}
\end{table} \FloatBarrier
\FloatBarrier \begin{table}[htbp]
\centering
\caption {The fifth mode shape plots for different ply orientations for $\gamma=10$ and $\Delta\epsilon=0$} \label{tab:vib6} 
\centering
\begin{tabular}{|c|c|}
\hline
  \textbf{\begin{tabular}[c]{@{}c@{}}Antisymmetric Crossply\end{tabular}} & \textbf{\begin{tabular}[c]{@{}c@{}}Symmetric Crossply\end{tabular}} \\ \hline  \hline
		\includegraphics[scale=.35]{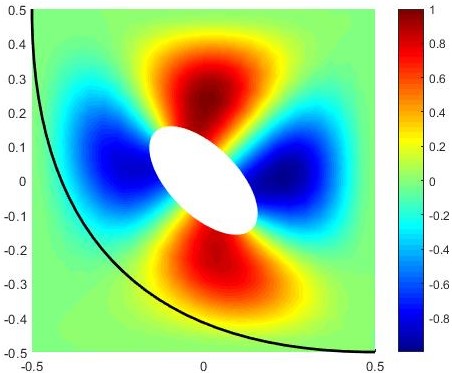}	&	\includegraphics[scale=.35]{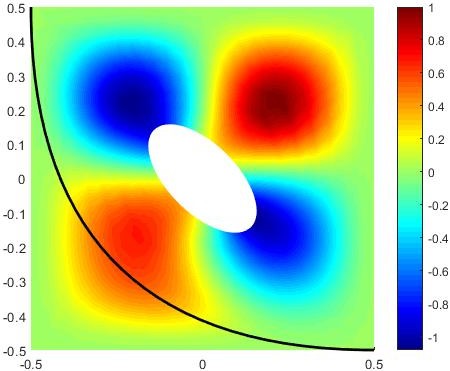}	\\ \hline
		\textbf{\begin{tabular}[c]{@{}c@{}}Symmetric Angleply\end{tabular}} & \textbf{\begin{tabular}[c]{@{}c@{}}Antisymmetric Angleply\end{tabular}} \\ \hline
		\includegraphics[scale=.35]{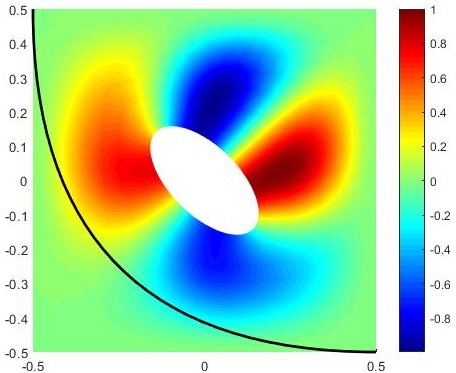}	&	\includegraphics[scale=.35]{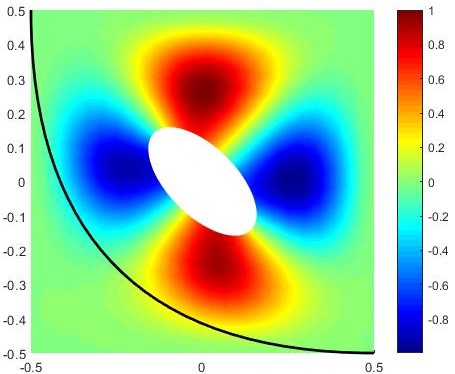}	\\	\hline
\end{tabular}
\end{table} \FloatBarrier
\FloatBarrier \begin{table}[htbp]
\centering
\caption {The fifth mode shape plots for different ply orientations for $\gamma=10$ and $\Delta\epsilon=0.25$} 
\label{tab:vib8} 
\centering
\begin{tabular}{|c|c|}
\hline
  \textbf{\begin{tabular}[c]{@{}c@{}}Antisymmetric Crossply\end{tabular}} & \textbf{\begin{tabular}[c]{@{}c@{}}Symmetric Crossply\end{tabular}} \\ \hline  \hline
		\includegraphics[scale=.35]{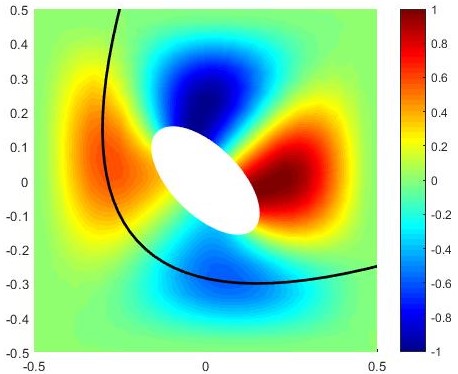}	&	\includegraphics[scale=.35]{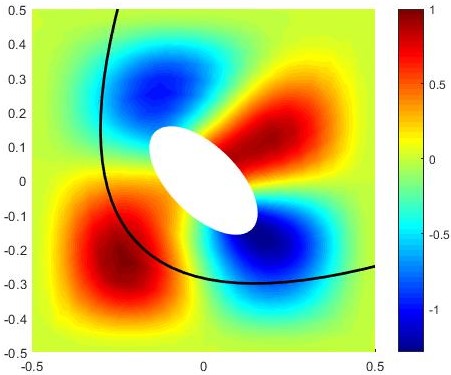}	\\ \hline
		\textbf{\begin{tabular}[c]{@{}c@{}}Symmetric Angleply\end{tabular}} & \textbf{\begin{tabular}[c]{@{}c@{}}Antisymmetric Angleply\end{tabular}} \\ \hline
		\includegraphics[scale=.35]{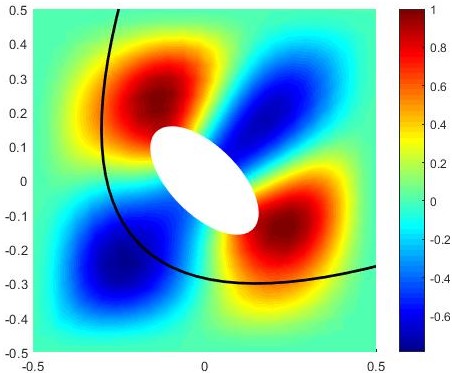}	&	\includegraphics[scale=.35]{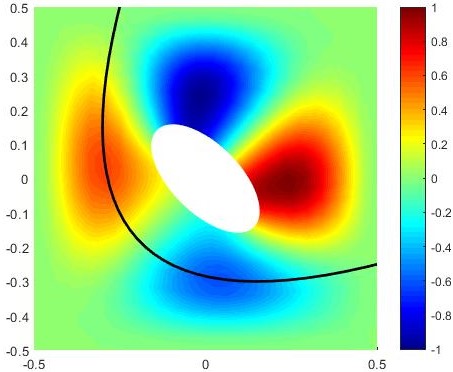}	\\	\hline
\end{tabular}
\end{table} \FloatBarrier
This section discusses the vibration analysis of curvilinearly stiffened plate with cutouts using a single NURBS patch. The efficiency of this method is described using numerical examples of laminated plates containing cutouts of various shapes and dimensions and for different boundary conditions, stiffener
locations and curvature.

\section{Conclusions}
\label{sec:sample:appendix}
The application of the isogeometric technique has proven effective in conducting vibration analysis on a curvilinearly stiffened plate containing cutouts. This was achieved by employing a sole NURBS patch. The method's effectiveness is demonstrated through numerical instances involving intricate cutout designs of various sizes and shapes, alongside diverse boundary conditions, locations of stiffeners, and curvatures for stiffened laminated plates. The utilization of a single NURBS patch to represent complex cutout shapes showcases impressive conformity between the outcomes obtained and those generated by contemporary methodologies. The isogeometric analysis a computationally efficient method that could be used along with powerful modern optimization algorithms like Particle Swarm Optimization and Differential Evolution ~\cite{biswas2021improving, saha2022chagskode, saha2022framework} to perform structural optimization and come up parameters for with light-weight designs.


 \bibliography{elsarticle-template-num}





\end{document}